\documentclass[pdftex,twocolumn,epjc3]{svjour3}          

\RequirePackage[T1]{fontenc}

\smartqed  

\RequirePackage{graphicx}
\RequirePackage{mathptmx}      
\RequirePackage{flushend}
\RequirePackage[numbers,sort&compress]{natbib}
\RequirePackage[colorlinks,citecolor=blue,urlcolor=blue,linkcolor=blue]{hyperref}

\usepackage{booktabs}
\usepackage{array}

\begin{document}

\title{Dynamic Vulnerability Criticality Calculator for Industrial Control Systems}

\author{Pavlos Cheimonidis \thanksref{addr1,e1}
        \and
        Kontantinos Rantos \thanksref{addr1,e2} 
}

\thankstext{e1}{e-mail: paxeimw@cs.ihu.gr}
\thankstext{e2}{e-mail: krantos@cs.ihu.gr}

\institute{Department of Computer Science, International Hellenic University, 654 04 Kavala, Greece \label{addr1}
}

\date{Received: date / Accepted: date}

\maketitle

\begin{abstract}
The convergence of information and communication technologies has introduced new and advanced capabilities to Industrial Control Systems. However, concurrently, it has heightened their vulnerability to cyber attacks. Consequently, the imperative for new security methods has emerged as a critical need for these organizations to effectively identify and mitigate potential threats. This paper introduces an innovative approach by proposing a dynamic vulnerability criticality calculator. Our methodology encompasses the analysis of environmental topology and the effectiveness of deployed security mechanisms,  coupled with the utilization of the Common Vulnerability Scoring System framework to adjust detected vulnerabilities based on the specific environment. Moreover, it evaluates the quantity of vulnerabilities and their interdependencies within each asset. Additionally, our approach integrates these factors into a comprehensive Fuzzy Cognitive Map model, incorporating attack paths to holistically assess the overall vulnerability score. To validate the efficacy of our proposed method, we present a relative case study alongside several modified scenarios, demonstrating its effectiveness in practical applications.
\end{abstract}

\section{Introduction}

Supervisory Control and Data Acquisition (SCADA) systems play
a vital role in overseeing and governing a wide spectrum of operations within various Industrial Control System (ICS) environments~\cite{TARIQ2019612}. At its inception, SCADA systems were conceived as autonomous frameworks designed to manage and control localized networks within ICS~\cite{5357008}. However, the landscape has evolved with the integration of Information and Communication Technologies (ICT), eroding the once-prevalent isolation. Consequently, ICS, including SCADA, are currently facing significant and increasing cybersecurity threats~\cite{TrautmanOrmerod2018}.

The extensive integration of ICT has empowered ICS to acquire enhanced capabilities, notably enabling connectivity with diverse networks. Simultaneously, this integration has exposed ICS to an increasing risk of infection by viruses, Trojan horses, and other pervasive threats~\cite {SonmezKilic2021}. Consequently, the (cyber) security challenge associated with ICS has become significantly more severe. Therefore, the assessment of risk within ICS becomes imperative, assuming a pivotal role in quickly detecting and addressing security issues in ICS operations~\cite {RenXuDaiZhang2021}.

In today's cybersecurity landscape, ICS cannot depend on static 
and inflexible risk assessment frameworks and processes. These frameworks are inadequate to maintain the security posture at the necessary
levels due to their incapacity to adapt effectively to
modern and dynamic security environments~\cite {linkov2014risk}.
Static risk assessment frameworks and methods conduct risk evaluations at predetermined intervals, lacking continuous assessment
capabilities. Consequently, this approach engenders misconceptions regarding threats and their potential impacts due to its inability to capture real-time changes and evolving risks~\cite{naumov2016dynamic}.
The necessity for such capabilities has spurred the development of Dynamic Risk Assessment (DRA) methods, which entail the ongoing process of identifying and evaluating risks associated with organizational operations in (or nearly in) real-time~\cite{cheimonidis_rantos_2023}. DRA is an invaluable tool capable of assessing and mitigating cyber threats within today's complex and rapidly evolving environments. It serves not only to evaluate and counter these threats in real-time, but also to elevate the security posture to the most appropriate level.

Our previous work, which focused on analyzing DRA methods, has revealed that a majority of these methods provide a comprehensive estimation framework primarily focused on evaluating changes in the probability or likelihood of threat occurrence~\cite{cheimonidis_rantos_2023}. This estimation is based on the premise that risk can be assessed as a result of probability, impact, and vulnerability. 

While acknowledging that dynamically updating one of these risk factors contributes to the overall risk value update, we strongly advocate for a DRA calculator capable of dynamically updating both vulnerability and probability values and, in certain scenarios, impact as well. The main aim of this study is to create a dynamic vulnerability calculator that concentrates on the ongoing assessment of vulnerability criticality within a specific environment. Our approach stands out from the prevailing DRA methodologies by not only adapting the vulnerability criticality value in response to relevant information unique to the target environment but also by considering the interdependencies among vulnerabilities. We also account for how modifications to one vulnerability value might impact another vulnerability value within the system. These adjustments allow our model to generate a dynamic vulnerability-only value.

Our model leverages the Common Vulnerabilities and Exposures (CVE) 
along with the Common Vulnerability Scoring System (CVSS) records to identify vulnerabilities within the assets of the target environment. Specifically, we utilize the identified CVEs  that are related to these assets to retrieve the corresponding CVSS vector strings and the relative score from the National Institute of Standards and Technology (NIST) National Vulnerability Database (NVD) website. The vector string is then employed by our model to retrieve exploitability-specific metrics. Intentionally, we disregard any metrics associated with Impact and Scope. We leverage exploitability metrics in conjunction with information regarding the environment's topology and implemented security measures. This amalgamation facilitates the creation of modified (adjusted) exploitability metrics which play a pivotal role in our methodology.

Our decision to adopt this approach stems from our aspiration in future endeavors to integrate the value derived from our model into a DRA model. This forthcoming model will encompass the multiplication of dynamic vulnerability criticality, dynamically assessed threat probability, and impact (which will not be included in the vulnerability-value as it is now the case with CVSS), culminating in the generation of a dynamic risk score.

The subsequent sections of this paper are organized as follows: section \ref{back} presents background information relevant to the subject matter addressed in this study,
section \ref{sec:related} delves into related work, section \ref{sec:methodology} details our methodology, and section \ref{sec:exp} encapsulates a case study, along with several scenarios, in which our approach is implemented, presenting the results and facilitating a detailed discussion. Finally, section \ref{sec:conc} offers conclusive remarks on our study and references future work.

\section{Background} \label{back}

The CVE~\cite{mitre}, established by MITRE, serves as a repository documenting individual vulnerabilities. It is widely recognized as the industry standard for numerous products. The CVSS~\cite{cvss} acts as an open framework integrating risk attributes to assess the severity of vulnerabilities. A CVSS vulnerability is represented as a vector string, which serves as a condensed textual representation encompassing the values utilized in deriving the score. 

The most recent version of CVSS, v.4.0, consists of four distinct metric groups: Base, Threat, Environmental, and Supplemental, each comprising a range of metrics. Given that there are currently no CVEs evaluated using the CVSS v.4.0 framework, this study will employ the 3.1 version of CVSS for assessing CVEs. In the CVSS 3.1 framework, metrics are categorized into three groups:
\begin {itemize}
    \item The Base Metric group delineates the inherent characteristics of a vulnerability, which remain constant across various environments and over time.
    \item  The Temporal Metric group assesses the present status of exploit availability or code readiness, including the existence of patches or workarounds.
    \item  The Environmental Metric group enables the customization of the CVSS score by considering the importance of affected IT assets in terms of Confidentiality, Integrity, and Availability(CIA). Furthermore, it incorporates the evaluation of existing security controls, facilitating a customized assessment based on environmental specifics.
\end{itemize}

The Base Group includes the  Exploitability, Impact and Scope Metrics. The Exploitability Metrics encompass the:
\begin{itemize}
    \item Attack Vector (AV) presents the context through which a vulnerability can successfully be exploited considering the remoteness of an attacker from a networking perspective, with potential values of Network (N), Adjacent (A), Local (L), and Physical (P);
    \item Attack Complexity (AC) assesses the difficulty an attacker faces in successfully exploiting the vulnerability with potential values of Low (L) or High (H);
    \item Privileges Required (PR) describes the level of privileges an attacker should have in order to successfully exploit the vulnerability with potential values of None (N), Low (L), or High (H); and
    \item User Interaction (UI) signifies the requirement for an arbitrary user, distinct from the attacker, to be present for the vulnerability to be successfully exploited, with potential values of None (N) or Required (R).
\end{itemize} 

In addition,  Impact Metrics cover Confidentiality (C), Integrity (I), and Availability (A), denoted on a scale of None, Low, or High. Furthermore, there is the  Scope (S) with values indicating Unchanged or Changed. The previously mentioned metrics of Exploitability, Impact, and Scope contribute to the derivation of the exploitability subscore, impact subscore, and scope subscore accordingly. These subscores are then amalgamated to yield an overall base score.

To effectively utilize the CVSS standard and conduct a reassessment of a vulnerability, the incorporation of the Environmental Metric Group is essential. The Environmental Metric Group encompasses:

\begin{itemize}
    \item \textbf{Exploitability Metrics:}
    \begin{itemize}
        \item Modified Attack Vector (MAV)
        \item Modified Attack Complexity (MAC)
        \item Modified Privileges Required (MPR)
        \item Modified User Interaction (MUI)
    \end{itemize}
    
    \item \textbf{Scope Metrics:}
    \begin{itemize}
        \item Modified Scope (MS)
    \end{itemize}
    
    \item \textbf{Impact Metrics:}
    \begin{itemize}
        \item Modified Confidentiality (MC)
        \item Modified Integrity (MI)
        \item Modified Availability (MA)
    \end{itemize}
    
    \item \textbf{Impact Subscore Modifiers Metric:}
    \begin{itemize}
        \item Confidentiality Requirement (CR)
        \item Integrity Requirement (IR)
        \item Availability Requirement (AR)
    \end{itemize}
\end{itemize}

The Environmental Metric Group includes all Base Metrics (Exploitability, Impact and Scope), albeit in modified formats, while also introducing supplementary modifiers — CR, IR, and AR. These aforementioned modifiers provide analysts with the ability to tailor the CVSS score, considering the significance of the CIA aspects of affected IT assets within a user's organization, relative to other impact factors. The CVSS 3.1 calculator generates three distinct values corresponding to the three discrete metric groups: Base, Temporal, and Environmental.

The NIST NVD~\cite{nvd} stands as the official repository within the U.S. government, hosting comprehensive data for standards-compliant vulnerability management. Within the NVD, scrutiny of CVEs listed in the CVE Dictionary is conducted. NVD's responsibility extends to the thorough analysis of CVEs, aggregating details derived from their descriptions, referenced sources, and any available supplementary public data during assessment. This process culminates in the determination of associated metrics employing the CVSS framework. In addition, the NVD provides an aggregate score based on the 3.1 framework, amalgamating all three distinct metric groups: Base, Temporal, and Environmental.

In instances where the Environmental Metric Group is not utilized, the score is contingent solely upon the metrics of the Base Metric Group. However, when adjustments are applied to metrics within the Environmental Metric Group, the modified metrics supersede the metrics of Base Metric Group, leading to the score being predominantly influenced by the environmental metrics.

The Common Attack Pattern Enumeration and Classification (CAPEC)~\cite{capec} initiative serves as an accessible repository, offering a comprehensive list of attack patterns. These patterns aid in comprehending how adversaries exploit vulnerabilities within applications and various cyber-enabled capabilities.
CAPEC encompasses detailed depictions of the recurring attributes and methodologies employed by adversaries to exploit identified weaknesses in cyber-enabled capabilities.

\section{Related work}\label{sec:related}

Zhu et al.~\cite{zhang2018fuzzy} introduced a model to quantitatively evaluate risk in Industrial Production Systems by assessing probabilities and consequences of abnormal events. This model integrates two components - probability inference and loss calculation -- to provide real-time risk assessments. Using an Extended Multilevel Flow Model (EMFM), it quantitatively describes the production process, aiding in forecasting consequences of abnormal events. Employing a Bayesian network based on the EMFM, it deduces the probabilities of abnormal events, incorporating established control strategies and evidence from Intrusion Detection System (IDS).

Li et al.~\cite{li2018asset} introduced a dynamic impact assessment method for ICSs. The approach abstracts assets, considering their properties, and utilizes function and performance attributes to categorize them into five groups based on their roles within the system. This process leads to the creation of component-level asset models, followed by system-level asset models that capture their interactive relationships. The system incorporates information from the IDS as input, integrating it into both component and system asset models for impact propagation analysis. Analyzing insights from asset, attack, and hazardous incident domains, it quantifies the impact, which, when combined with impact propagation analysis, yields the total impact.

Peng et al.~\cite{peng2018modeldata} introduced a method for quantifying cyber risk in ICS environments. Real-time attack evidence is integrated into a Bayesian network alongside with ICS security knowledge encompassing vulnerabilities, functions, accidents, and assets. The Bayesian network output provides the event occurrence probability, combined with impact assessment to derive real-time risk.

Huang et al.~\cite{huang2017application} introduced a dynamic model based on Bayesian networks to handle cyber risk in SCADA environments. The model utilizes Bayesian networks to combine posterior probability and asset value within the SCADA system to determine risk levels. It segregates nodes into vulnerability nodes (with detectable vulnerabilities) and privilege escalation nodes (with potential system damage capabilities). Real-time data collected by the IDS drives the model to predict posterior probabilities. By amalgamating historical data and real-time observations, this model leverages machine learning techniques to enhance the accuracy of risk estimations.

Qin et al.~\cite{qin2021association} proposed a dynamic cyber risk model for ICS using association analysis. The model comprises two core components: probability inference and risk quantification. Initially, system knowledge and historical security data construct an association network for probability inference. In parallel, the model creates an association matrix by mining state variables of crucial ICT assets in an offline stage. Subsequently, real-time IDS data feed into the probability inference for attack probability predictions during the online stage. The output integrates with the association matrix to compute the system's risk.

Wu et al.~\cite{wu2018security} proposed a security assessment method for ICS employing an ontology and an attack graph. The system's elements - assets, vulnerabilities, attacks, counter-measures, and their interconnections - are structured within an ontology using Web Ontology Language (OWL). This security ontology identifies attacks that pose threats to assets based on their vulnerabilities, facilitated by Semantic Web Rule Language (SWRL) rules. The attack graph is then generated utilizing an efficient algorithm that leverages the capabilities embedded within the security ontology.

Yan et al.~\cite{yan2023cyber} introduced a DRA  model tailored for SCADA networks, specifically evaluating the quantitative physical repercussions stemming from cyber attacks within these systems. This model incorporates CVSS scoring to gauge the severity and exploit likelihood of software vulnerabilities. It considers various factors including time, attacker characteristics, network security defenses, and propagation dynamics, utilizing mathematical formulations to assess the likelihood of exploitation. 

Vasilyev et al.~\cite{vasilyev_cybersecurity_2021} introduced a cybersecurity risk assessment model designed specifically for ICS environments. This model integrates existing vulnerabilities and CAPEC to formulate potential attack pathways. Subsequently, they employ Fuzzy Cognitive Map (FCM) in conjunction with the CVSS metric to compute the overall risk exposure within the system.

The majority of the previously mentioned studies concentrate on establishing a dynamic (cyber) risk score, primarily centered on altering the probability of threat occurrence factor. Many of these models employ the Bayesian networks method which is a  probabilistic approach. One  model ~\cite{li2018asset} offers a dynamic impact score calculation, but none of them provides a dynamic vulnerability criticality assessment similar to our proposed approach. The work of Vasilyev et al.~\cite{vasilyev_cybersecurity_2021} shares similarities with ours; however, the following distinctions are noteworthy:

\begin{enumerate}
    \item Vasilyev's model offers insights into the impact metrics derived from the CVSS framework, delineating alterations in these metrics when implementing CVSS within an ICS environment as compared to an ICT environment. We, conversely, specifically elaborate on the modified exploitability metrics, as our model aims to provide a dynamic vulnerability-only value. Therefore, our focus excludes the impact and scope metrics.
    \item Our modification of the exploitability metrics is contingent on the target environment's topology and the deployed security mechanisms.
    \item We adopt the attack tree framework to account for situations where an asset might have multiple vulnerabilities, but also, consider their interdependencies.
    \item Multiple scenarios are presented, showcasing the implementation of our methodology, to offer valuable insights into the effectiveness of measures aimed at diminishing the vulnerability score.
\end{enumerate}

\section{The Proposed Vulnerability Criticality Assessment Methodology} \label{sec:methodology}

\begin{figure}[ht]
\centering
\includegraphics[width=\columnwidth]{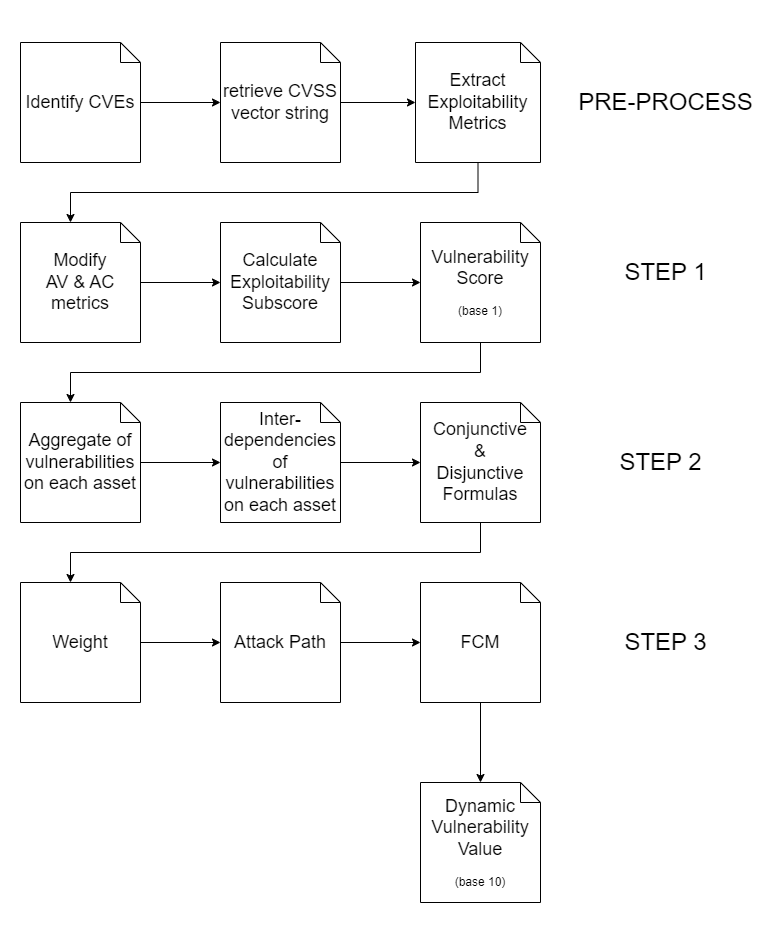}
\caption{Methodology's flowchart}
\label{FWL}
\end{figure}

To expound on our methodology, the pre-process step involves the identification of all CVEs, in order to retrieve the CVSS vector strings and then extract the exploitability metrics. The next step involves adjusting the AV and AC metrics of assets based on the environment's topology and the applied security measures accordingly, in order to  recalculate the exploitability subscore, henceforth referred to as the exploitability score. Subsequently, we normalize this score to a 1-point scale and rename it as vulnerability score. Following this normalization, we analyze the aggregate of identified vulnerabilities (CVEs) within each asset, considering their interdependencies (AND/OR) via the attack tree formulas. This process yields an aggregate vulnerability score for each asset, which is utilized as a weight within our FCM model. In the final step of our methodology, we analyze the assets' topology along with their vulnerabilities to generate potential attack paths. These paths are integrated into our FCM model, accompanied by the estimated weights. Upon executing and stabilizing the FCM model, the values which range from 0 to 1, are converted into a base-10 system, providing the final output of our methodology, which is the Dynamic Vulnerability Value. Figure~\ref{FWL} presents the flowchart of our methodology.

\subsection{Adjustments to base scores - Step 1} \label{sec:adj}

Our model employs CVEs alongside their respective CVSS vector strings, to extract exploitability metrics. The exploitability score can be calculated using formula (\ref{explot}) \cite{nvd, wang_cvss-based_2020, vilches2018towards, kurniawan_automation_2023, ur-rehman_vulnerability_2019}: 

\begin{equation} \label{explot}
    E = 8.22 * AV * AC * PR * UI 
\end{equation}

The 8.22 constant, originating from the NVD exploitability score equation, serves as a weighting factor influencing the effect of metrics contributing to the exploitability score. Table~\ref{tab:cvss_values} delineates the aforementioned metrics employed within the exploitability equation of the CVSS 3.1 framework, demonstrating their corresponding potential values~\cite{purkayastha_continuous_2020}. Considering the exploitability equation and the potential values outlined in Table \ref{tab:cvss_values}, the maximum exploitability score achievable is calculated at 3.9.

\begin{table*}
\caption{CVSS 3.1 exploitability metrics and values}
\label{tab:cvss_values}
\begin{tabular*}{\textwidth}{@{\extracolsep{\fill}}lrrr@{}}
\hline
\textbf{Metric} & \multicolumn{1}{c}{\textbf{Options}} & \multicolumn{1}{c}{\textbf{Value}} \\
\hline
Attack Vector(AV) - Modified Attack Vector (MAV) & Network (N) & 0.85 \\
 & Adjacent (A) & 0.62 \\
 & Local (L) & 0.55 \\
 & Physical (P) & 0.20 \\
\hline
Attack Complexity (AC) - Modified Attack Complexity (MAC) & Low (L) & 0.77 \\
 & High (H) & 0.44 \\
\hline
Privileges Required (PR) - Modified Privileges Required (MPR) & None (N) & 0.85 \\
 & Low (L) & 0.62 \\
 & High (H) & 0.27 \\
\hline
User Interaction (UI) - Modified User Interaction (MUI) & None (N) & 0.85 \\
 & Required (R) & 0.62 \\
\hline
\end{tabular*}
\end{table*}

As per the guidelines within the CVSS framework, making adjustments to a CVSS score entails modifying the values within the environmental metric group rather than altering the base metric group. For our study, we utilize the modified exploitability metrics. Utilizing the  equation~\ref{mexplot} \cite{nvd, vilches2018towards}, we can compute the modified exploitability score by incorporating the modified metrics and their respective values. The modified values correspond to those provided in Table \ref{tab:cvss_values}, and hence the maximum modified exploitability score denoted as ME is also 3.9.

\begin{equation} \label{mexplot}
   ME = 8.22 * MAV * MAC * MPR * MUI  
\end{equation}

To illustrate this concept, consider the following example: suppose we have a vulnerability represented by the above string: 

\begin{equation} 
   AV:N, AC:L, PR:N, UI:N  
\end{equation}

This configuration indicates that the vulnerability can be exploited through a remote or distant network (\begin{math}AV:N\end{math}). Moreover, the (\begin{math}AC:L\end{math}) associated with exploiting this vulnerability is relatively low, implying that the complexity of launching an attack is not highly intricate, implying fewer prerequisites or security measures to a successful attack. Additionally, it requires no special privileges (\begin{math}PR:N \end{math}) to exploit, meaning that even without elevated access rights, an attacker could potentially exploit the vulnerability. Furthermore, the absence of user interaction (\begin{math}UI:N
\end{math}) signifies that an attacker can carry out the exploit without needing any action or input from a legitimate user. This configuration results in an exploitability score of 3.9, according to NVD.

Our model incorporates the topology and existing security mechanisms to tailor vulnerability scores, ensuring alignment with the specific environment rather than adhering to generic assessments. More specifically, our model comprehensively considers and adjusts the AV metric based on the target environment's topology and the AC metric considering implemented security mechanisms. The AV metric delineates the context through which a vulnerability may be exploited, based on how remote an attacker can be, from a networking perspective, to the targeted vulnerable system ~\cite{franklin_cvss_2014}. A higher metric corresponds to situations where an attacker can exploit a vulnerability from remote networks. Meanwhile, the AC metric refers to an assessment of the conditions beyond the attacker's control that must exist to exploit a vulnerability. It evaluates the level of difficulty in launching an attack and the prerequisites necessary for a successful exploit. The level of the AC metric within the context of the CVSS can be influenced by the presence or absence of security controls~\cite {stellios_assessing_2021}. 

To highlight the alterations made by our model, consider the same example as previously used:

\begin{equation} 
   AV:N, AC:L, PR:N, UI:N  
\end{equation}

Now, we examine the scenario in which this asset is inaccessible from external networks, including the internet, and is fortified by a robust security mechanism. This configuration transforms the string to: 

\begin{equation} 
   MAV:A, MAC:L, MPR:N, MUI:N  
\end{equation}

The calculation of the adjusted exploitability score, resulting in a value of 1.6, signifies a notable reduction in the overall vulnerability criticality associated with the specific asset. This underscores the importance of considering the target's network topology, evaluating the asset's susceptibility to attacks, and acknowledging the pivotal role of security mechanisms deployed to safeguard assets within our environment. Finally, this highlights the importance of dynamically calculating vulnerability criticality, incorporating all aforementioned features specific to the environment in which a vulnerability is identified. The dynamic approach recognizes the unique attributes of each environment, including network topology and security mechanisms, enabling a more precise and context-aware assessment of vulnerability criticality. Emphasizing the avoidance of reliance on generic assumptions, which may lead to misleading conclusions, this nuanced evaluation facilitates a comprehensive understanding of the detected vulnerabilities. 

In examining the environmental topology, our assessment involves evaluating the reachability of assets' vulnerabilities while considering the original AV metric from the attacker's perspective. If Asset X possesses (\begin{math}AV:N\end{math}), and its vulnerability can be exploited by assets situated in remote networks, the MAV remains N. However, if Asset X has (\begin{math}AV:N\end{math})  but, within the examined topology, is exploitable solely by assets on the same logical network, we designate the MAV as A. In scenarios where the AV is not (N) but categorized as (A), (L), or (P), the MAV reflects the AV, indicating that the vulnerability's exploitation is already restricted in terms of attacker accessibility. To elaborate, if a vulnerability is labeled (\begin{math}AV:P\end{math}), yet accessible within our topology by another asset on the same network, we do not designate (\begin{math}MAV:A\end{math}). This decision is based on the premise that for this vulnerability to be exploited, the attacker requires Physical (P) access. This principle similarly applies to instances of (\begin{math}AV:A\end{math}) and (\begin{math}AV:L\end{math}). All the above principles collectively establish the first rule in our model (RULE\#1), which is used to calculate the MAV metrics based on AV metrics and environment's topology, as detailed in Table \ref{tab:AV}.

\begin{table}[htbp]
\centering
\caption{RULE\#1 - Calculation of MAV metrics}
\label{tab:AV}
\begin{tabular*}{\columnwidth}{@{\extracolsep{\fill}}ccc@{}}
\hline
\textbf{AV} & \textbf{Reachability} & \textbf{MAV} \\
\hline
N & Remote & N \\
N & Logical & A \\
A & N/A\textsuperscript{1} & A \\
L & N/A & L \\
P & N/A & P \\
\hline
\end{tabular*}
\\
\footnotesize{\textsuperscript{1} N/A: Not Applicable.}
\end{table}

Furthermore, our methodology evaluates the implemented security measures in conjunction with the AC metric from the attacker's perspective. In the scenario where Asset X has (\begin{math}AC:L\end{math}) and lacks any security protection, the MAC remains at L. If Asset X has (\begin{math}AC:L\end{math}) but is protected by a security mechanism, we designate MAC as H. Nevertheless, if AC is already rated as H, the MAC maintains its H value irrespective of the existence of a security mechanism. This occurrence transpires because (\begin{math}AC:H\end{math}) implies that the attacker is already required to bypass intrinsic security measures.
All the aforementioned principles collectively constitute the second rule of our model (RULE\#2), employed to calculate the MAC metrics based on the AC metrics and the presence or absence of applied security mechanisms, as outlined in Table \ref{tab:AC}.

\begin{table}[htbp]
    \centering
    \caption{RULE\#2 - Calculation of MAC metrics}
    \label{tab:AC}
    \begin{tabular*}{\columnwidth}{@{\extracolsep{\fill}}ccc@{}}
        \toprule
        \textbf{AC} & \textbf{Security Mechanism} & \textbf{MAC} \\
        \midrule
        L & YES & H \\
        L & NO & L \\
        H & N/A & H \\
        \bottomrule
    \end{tabular*}
\end{table}

To ensure a comprehensive outcome, which can be used as input at the next step of our methodology we opt to re-scale the 3.9 exploitability score to a base of 1. Consequently, the maximum exploitability score assignable to a CVE within our model will be adjusted from 3.9 to 1. All mentioned CVEs' scores are recalculated based on this adjustment.

Furthermore, given our model's exclusive focus on the exploitability metric, the CVSS score of this metric, henceforth set at 1, will be referred to as the vulnerability score rather than the exploitability score. It's important to note that the other two exploitability metrics (Privileges Required -- PR and User Interaction -- UI) remain unchanged in our model.

\subsection{Attack Trees - Step 2} \label{AT}

Attack trees represent a method for modeling the security threats within information systems. They delineate various pathways an adversary can follow to attain their objective, typically reaching the final target. These trees construct a hierarchical structure based on these pathways, with elementary attacks positioned at the leaf level and the principal attack situated at the root~\cite{haque2017evolutionary}. In this approach, the nodes are perceived as expansions or elaborations of higher-level nodes and can exist as either conjunctive (aggregation or 'AND' nodes) or disjunctive (choice or 'OR' nodes)~\cite{petrica_studying_2017}.

Conjunctive relationships can be represented by formula (\ref{AND}):
\begin{equation} \label{AND}
 Pc = \prod_{i=1}^{n} P(i) 
\end{equation}

Disjunctive relationships can be represented by formula (\ref{OReq}):
\begin{equation} \label{OReq}
Pd = 1 - \prod_{i=1}^{n} \left(1-P_{(i)}\right) 
\end{equation}

\begin{figure}[ht]
\centering
\includegraphics[width=\columnwidth]{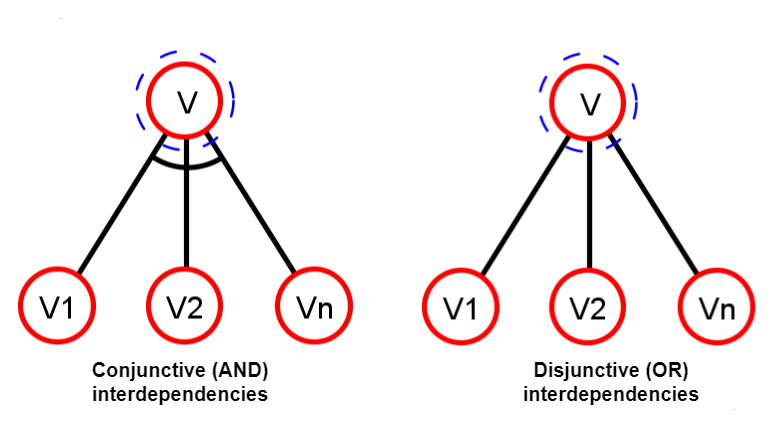}
\caption{Conjunctive and disjunctive relationships}
\label{OR}
\end{figure}

In our methodology, we leverage both conjunctive and disjunctive options to comprehensively evaluate all vulnerabilities associated with a single node. This approach enables the recalibration of a node's overall vulnerability score by accounting for the number of vulnerabilities present on each node, along with their respective interdependencies (conjunctive, disjunctive). Figure~\ref{OR} illustrates the aggregate of vulnerabilities (V1, V2,..., Vn) using conjunctive and disjunctive approaches accordingly~\cite{adtool}.

In our approach, when a node contains multiple vulnerabilities that result in the same or similar impact from the attacker's perspective, we utilize the disjunctive formula. This is based on the assumption that the attacker needs to exploit only one of these vulnerabilities to achieve their objective. For instance, the successful exploitation of vulnerabilities V1, V2,..., Vn enables the attacker to remotely execute code. The integration of the aforementioned principle establish the (RULE\#3) in our model.  Conversely, if a node possesses multiple vulnerabilities that require exploitation, we utilize the conjunctive formula. This decision is based on the assumption that the attacker needs to exploit all of these vulnerabilities to accomplish their objective. This principle constitutes the concluding rule in our model, denoted as (RULE\#4).

\subsection{Fuzzy Cognitive Maps - Step 3}

FCMs have found extensive application in the analysis of intricate systems driven by causality. These applications encompass modeling, decision-making, analysis, prediction~\cite{bakhtavar_et_al_2021} and cybersecurity~\cite{fan_et_al_2021}. FCMs offer a robust approach for modeling relationships among various factors within a system. They have a lot of characteristics, some of which are related to our model: 

\begin{enumerate}
\item  Signed Causality: FCMs incorporate signed causality to denote positive or negative relationships, elucidating both the direction and nature of causality between factors~\cite{papageorgiou_stylios_2008}.
\item  Strength or Weigh of Causal Relationships: these models embrace fuzzy values to represent the strengths/weights of causal relationships. This characteristic assigns fuzzy numbers or values which range from [-1,1], delineating the intensity or degree of association between two concepts~\cite{napoles2018fcm}.
\item  Equilibrium Point: The concept of an equilibrium point in FCM refers to a state where the values of all interconnected nodes within the map no longer undergo changes across subsequent iterations~\cite{4801671}.
\end{enumerate}

In our methodology, we utilize FCM to elucidate the dependencies among vulnerabilities within the attack path, specifically examining how the vulnerability score of asset X influences the vulnerability score of asset Y. To elaborate:
\begin{enumerate}
    \item The signed causality within our model signifies the directional sequence of movements that an attacker could potentially undertake to transition from one asset to another, exploiting vulnerabilities along the path.
    \item The assigned weights denote the vulnerability score of each node, as derived from the analysis conducted in steps 1 and 2
    \item The equilibrium point offers insights into the dynamic vulnerability value at the final node. It signifies the state when our model achieves stability, presenting the overall vulnerability criticality assessment.
\end{enumerate}

\section{Case Study}\label{sec:exp}
In this study, we present an attack scenario targeting ICS by leveraging CAPEC tailored specifically for ICS. The purpose of utilizing the CAPEC framework is to construct a more realistic scenario that illustrates potential actions from an adversary's perspective within an ICS environment. In a detailed breakdown, the assumed attack sequence begins with footprinting, wherein the attacker aims to gather pertinent information about the targeted ICS environment. Subsequently, the attacker seeks to gain access to the system by targeting assets that maintain connectivity with the internet, such as the VPN server (VPN) and the Web server (WebS).

Following initial access, the attacker progresses to conduct network mapping, specifically identifying administrative WorkStations (WS). Subsequently, attempts are made to gain control of these workstations as they serve as pivotal access points capable of communicating and managing assets within the industrial layer of the network. Upon success, the attacker proceeds with active scanning activities to discern system-specific details, including operating systems and firmware information pertaining to the assets in the industrial layer.

Building on this new reconnaissance phase, the adversary aims to exploit remote services to gain unauthorized access to assets within the industrial layer. Having established unauthorized access, the adversary's objective shifts towards manipulating components in the field layer, particularly targeting the Programmable Logic Controller (PLC) to disrupt the operation of critical elements, such as the valve, thus causing a detrimental impact on the operational functionality of the ICS. Table \ref{tab:capec} displays the specific CAPEC Tactics utilized within the context of our scenario, providing a comprehensive overview of the relative attack patterns employed.

\begin{table*}
\caption{CAPEC descriptions}
\label{tab:capec}
\begin{tabular*}{\textwidth}{@{\extracolsep{\fill}}p{2.7cm}p{9cm}@{}}
\hline
\textbf{CAPEC Tactic} & \textbf{Description} \\
\hline
CAPEC-169 & An adversary engages in probing and exploration activities to identify constituents and properties of the target. \\

CAPEC-560 & An adversary guesses or obtains (i.e., steals or purchases) legitimate credentials (e.g., userID/password) to achieve authentication and to perform authorized actions under the guise of an authenticated user or service. \\

CAPEC-309 & An adversary engages in scanning activities to map network nodes, hosts, devices, and routes. \\

CAPEC-70 & An adversary may try certain common or default usernames and passwords to gain access into the system and perform unauthorized actions. \\

CAPEC-312 & An adversary engages in activity to detect the operating system or firmware version of a remote target by interrogating a device, server, or platform with a probe designed to solicit behavior that will reveal information about the operating systems or firmware in the environment. \\

CAPEC-555 & This pattern of attack involves an adversary that uses stolen credentials to leverage remote services such as RDP, telnet, SSH, and VNC to log into a system. Once access is gained, any number of malicious activities could be performed. \\

CAPEC-441 & An adversary installs or adds malicious logic (also known as malware) into a seemingly benign component of a fielded system. This logic is often hidden from the user of the system and works behind the scenes to achieve negative impacts. \\
\hline
\end{tabular*}
\end{table*}

It is important to note that in our specific scenario, we have opted to present a generalized description encompassing potential attack patterns that could compromise our environment and adversely impact the PLC. The selected CVEs, although not precisely aligned with the specific patterns outlined in the corresponding CAPEC entries referenced in Table \ref{tab:capec}, serve a purpose in demonstrating the efficacy of our methodology. This divergence emerges from our strategy to incorporate a comprehensive range of CVEs that specifically match realistic SCADA assets situated in the industrial layer within our case study. This includes CVEs pertinent to the Human Machine Interface (HMI), Engineering WorkStation (EWS), and PLC.

\subsection{Sample SCADA network}

SCADA networks encompass a wide array of devices, including WS, Historical Databases (HDB), HMI, EWS, PLC ~\cite{chandia2008security}. In our specific scenario, the SCADA network is structured into three foundational layers~\cite{wang2010simulation}: the enterprise layer, the industrial layer, and the field or physical layer. Each layer comprises distinct assets, some of which face susceptibility to multiple vulnerabilities.

\begin{figure}
\centering
\includegraphics[width=\columnwidth]{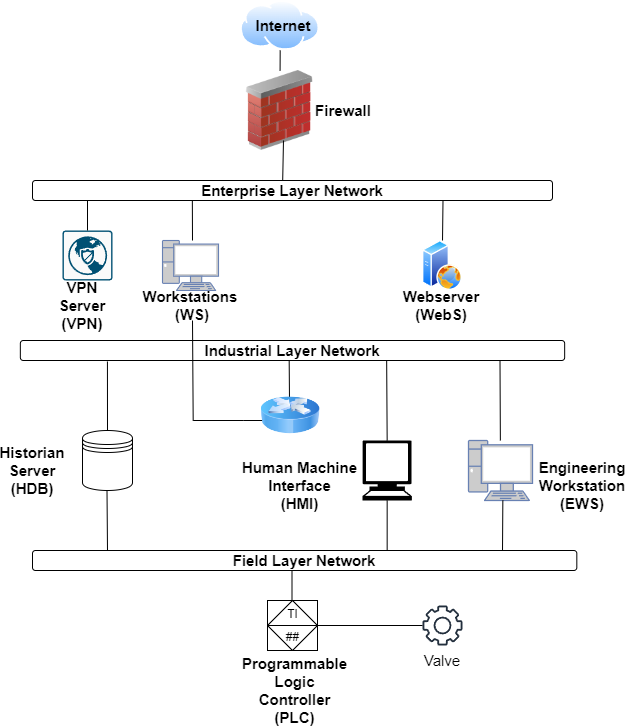}
\caption{Experimental environment}
\label{Env}
\end{figure}

Delving deeper, the cyber layer is protected with a firewall and consists of key components, such as the VPN utilized for remote connections, the WebS responsible for managing the company's website, and WS that communicate through a router with assets that are part of the industrial layer. Assets within the industrial layer encompass the HDB, entrusted with the storage and logging of all SCADA-related data, as well as the HMI and EWS, crucial for the supervision and management of the SCADA infrastructure. Lastly, the PLC, situated within the field layer, assumes operational control of the valve. Figure \ref{Env} illustrates the schematic representation of our environment, depicting the layers, assets, and their dependencies within the system.

\subsection{Methodology: Pre-Process - Vulnerability Analysis}

The pre-process step within our proposed vulnerability criticality assessment methodology necessitates identifying vulnerabilities within each asset of our SCADA network. This involves identifying CVEs and retrieving their respective CVSS vector strings to extract only the exploitability metrics. While the scenario employed for demonstrative purposes is fictitious, it endeavors to portray a realistic environment. 

In our scenario, the VPN is susceptible to CVE-2019-11510, allowing an unauthenticated remote attacker to exploit a specially crafted URI, thereby facilitating unauthorized file reading which could be used for authentication. Meanwhile, the WebS presents vulnerability CVE-2017-7269, permitting remote attackers to execute arbitrary code. Additionally, the WS has vulnerabilities CVE-2017-0143 and CVE-2017-8692, both of which provide opportunities for remote code execution.

In the context of the industrial layer, specific CVEs, namely CVE-2021-1636 and CVE-2023-21528, have been detected within the HDB. These vulnerabilities may be used by the attackers to gain access and then to execute code. Furthermore, the HMI is vulnerable due to CVE-2016-5743, enabling remote attackers to execute arbitrary code by leveraging crafted packets. Similarly, the EWS within this layer is susceptible to CVE-2019-10922, which also permits remote attackers to execute arbitrary code.

In the last layer of our SCADA network, the physical layer, the PLC is vulnerable to CVEs: CVE-2016-9159 and CVE-2016-8673. These vulnerabilities provide remote attackers the capability to access the PLC's credentials or compromise the authentication of arbitrary users.

Table \ref{tab:vul} presents a summary of the identified vulnerabilities categorized by the respective layer and associated assets, detailing the CVEs'-id and their original CVSS scores based on CVSS version 3.1. All information has been obtained from NIST NVD.

\begin{table}
\centering
\caption{Detected CVEs and their scores}
\label{tab:vul}
\begin{tabular*}{\columnwidth}{@{\extracolsep{\fill}}ccccc@{}}
\hline
\textbf{No} & \textbf{Layer} & \textbf{Host} & \textbf{CVE} & \textbf{CVSS Sc}  \\
\hline
V1 & Cyber & VPN & CVE-2019-11510 & 10  \\ 
V2 & Cyber & WebS & CVE-2017-7269 & 9.8   \\ 
V3 & Cyber & WS & CVE-2017-0143 & 8.1 \\
V4 & Cyber & WS & CVE-2017-8692 & 7.5 \\ 
V5 & Industrial & HDB & CVE-2021-1636 & 8.8 \\ 
V6 & Industrial & HDB & CVE-2023-21528 & 7.8 \\ 
V7 & Industrial & HMI & CVE-2016-5743 & 9.8 \\ 
V8 & Industrial & EWS &  CVE-2019-10922 & 9.8 \\
V9 & Physical & PLC & CVE-2016-9159 & 5.9 \\ 
V10 & Physical & PLC & CVE-2016-8673 & 8.8 \\ 
\hline
\end{tabular*}
\end{table}

As previously discussed in Section \ref{sec:adj}, the pre-process phase of our methodology centers on the extraction of exploitability metrics from the complete CVSS vector string. This extraction process aims to isolate and retain solely the exploitability metrics and score for further analysis. Table \ref{tab:exp} displays the original CVSS scores, the exploitability-only metrics along with the exploitability scores (not normalized) corresponding to the vulnerabilities utilized within our scenario. 

\begin{table*}
\centering
\caption{Detected CVEs and exploitability scores}
\label{tab:exp}
\begin{tabular*}{\textwidth}{@{\extracolsep{\fill}}cccc@{}}
\toprule
\textbf{CVE} & \textbf{CVSS score} & \textbf{Exploitability Metric} & \textbf{Exploitability score} \\ 
\midrule
CVE-2019-11510 & 10 & AV:N/AC:L/PR:N/UI:N & 3.9 \\ 
CVE-2017-7269 & 9.8 & AV:N/AC:L/PR:N/UI:N & 3.9 \\ 
CVE-2017-0143 & 8.1 & AV:N/AC:H/PR:N/UI:N & 2.2 \\ 
CVE-2017-8692 & 7.5 & AV:N/AC:H/PR:N/UI:R & 1.6 \\ 
CVE-2021-1636 & 8.8 & AV:N/AC:L/PR:L/UI:N & 2.8 \\ 
CVE-2023-21528 & 7.8 & AV:L/AC:L/PR:L/UI:N & 1.8 \\ 
CVE-2016-5743 & 9.8 & AV:N/AC:L/PR:N/UI:N & 3.9 \\ 
CVE-2019-10922 & 9.8 & AV:N/AC:L/PR:N/UI:N & 3.9\\ 
CVE-2016-9159 & 5.9 & AV:N/AC:H/PR:N/UI:N & 2.2 \\ 
CVE-2016-8673 & 8.8 & AV:N/AC:L/PR:N/UI:R& 2.8 \\ 
\bottomrule
\end{tabular*}
\end{table*}

\subsection{Methodology: Step 1 - Adjustments to base scores}

Continuing our methodology, the next step involves scrutinizing the topology and security mechanisms implemented to fine-tune the exploitability metrics (AV and AC) related to the aforementioned vulnerabilities. To elaborate further, we apply (RULE\#1) and (RULE\#2) of our methodology to ascertain the modified AV and AC accordingly.

As depicted in Figure \ref{Env}, the cyber layer is protected by a firewall. This security measure affects the AC metric for the vulnerabilities found in two specific assets, namely the WebS and the VPN, which have internet accessibility. Subsequently, our investigation commences with an assessment of the original AC metric within the exploitability vector string associated with these two assets. The subsequent step involves the application of (RULE\#2) from our methodology to determine the modified AC metric. Table \ref{tab:AdjV} illustrates the aforementioned assets alongside their original and modified exploitability vector strings. Upon examination of this table, it becomes evident that in both instances, the AC metric within the exploitability vector string has transitioned from `Low' (L) to `High' (H). This transition is a direct consequence of the implementation of (RULE\#2) in this specific scenario.

\begin{table*}
\centering
\caption{Modified exploitability (AC) metrics - (RULE\#2)}
\label{tab:AdjV}
\begin{tabular*}{\textwidth}{@{\extracolsep{\fill}}cccc@{}}
\hline
\textbf{Asset} & \textbf{CVE} & \textbf{Original Expl. String} & \textbf{Modified Exp. String} \\ \hline
VPN & CVE-2019-11510 & AV:N/AC:L/PR:N/UI:N & MAV:N/MAC:H/MPR:N/MUI:N \\ 
WebS & CVE-2017-7269 & AV:N/AC:L/PR:N/UI:N & MAV:N/MAC:H/MPR:N/MUI:N \\ \hline
\end{tabular*}
\end{table*}

As a result of the modified exploitability vector string, this alteration subsequently affects the modified exploitability score. Table \ref{tab:AdjS} outlines the comparison between the original and modified exploitability scores. Notably, the presence of the firewall causing an increase in the MAC metric has led to a decrease in the overall exploitability scores.

\begin{table*}
\centering
\caption{Modified exploitability (AC) scores}
\label{tab:AdjS}
\begin{tabular*}{\textwidth}{@{\extracolsep{\fill}}cccc@{}}
\hline
\textbf{Asset} & \textbf{CVE} & \textbf{Original Expl. Score} & \textbf{Modified Exp. Score} \\ \hline
VPN & CVE-2019-11510 & 3.9 & 2.2 \\ 
WebS & CVE-2017-7269 & 3.9 & 2.2 \\ \hline
\end{tabular*}
\end{table*}

The subsequent assets being examined within our methodology are the WS positioned in the cyber layer. As per the topology and scenario descriptions, these WS remain exclusively accessible from the internal network. This restricted accessibility significantly influences the `reachability' of these workstations, prompting a detailed assessment of the original AV metric within the exploitability vector string.

In our scenario, the attacker aims to breach our SCADA network by exploiting vulnerabilities present either in the VPN or the WebS accordingly. The WS, which communicates via the router with the distinct network of the industrial layer, cannot maintain the (\begin{math}AV:N\end{math}). This is due to the fact that for an adversary to gain access to these workstations, the intrusion must take place within the network of the cyber layer where the workstations are located. The adversary is unable to breach the network of the industrial layer without first infiltrating the network of the cyber layer, specifically targeting the WS. According to this presumption and the guidelines outlined in Table \ref{tab:AV}, when assets within our topology possess original (\begin{math}AV:N\end{math}), but are exploitable solely by assets on the same logical network, the resulting MAV is designated as A. This categorization is a direct application of (RULE\#1) within our methodology. Table \ref{tab:AdjV2} displays the modified exploitability vector strings for the WS.

\begin{table*}
\caption{Modified exploitability (AV) metrics - (RULE\#1)}
\label{tab:AdjV2}
\begin{tabular*}{\textwidth}{@{\extracolsep{\fill}}cccc@{}}
\hline
\textbf{Asset} & \textbf{CVE} & \textbf{Original Expl. String} & \textbf{Adjusted Exp. String} \\ \hline
WS & CVE-2017-0143 & AV:N/AC:H/PR:N/UI:N & MAV:A/MAC:H/MPR:N/MUI:N \\ 
WS & CVE-2017-8692 & AV:N/AC:H/PR:N/UI:R & MAV:A/MAC:H/MPR:N/MUI:R \\ \hline
\end{tabular*}
\end{table*}

As a consequence of the modified exploitability metrics, this adjustment consequently affects the exploitability score. Table \ref{tab:AdjS2} delineates the comparison between the original and modified exploitability scores for the WS. It's noteworthy that the specific topology dependencies and communication rules directly influence the MAV. In our scenario, the reduction in the exploitability score is a direct outcome of the alteration from (\begin{math}AV:N\end{math}) to (\begin{math}MAV:A\end{math}).

\begin{table*}
\caption{Modified exploitability (AV) scores}
\label{tab:AdjS2}
\begin{tabular*}{\textwidth}{@{\extracolsep{\fill}}cccc@{}}
\hline
\textbf{Asset} & \textbf{CVE} & \textbf{Original Expl. Score} & \textbf{Adjusted Exp. Score} \\ \hline
WS & CVE-2017-0143 & 2.2 & 1.6 \\ 
WS & CVE-2017-8692 & 1.6 & 1.2 \\ \hline
\end{tabular*}
\end{table*}

According to our scenario, the vulnerabilities and their corresponding exploitability metrics for the remaining assets remain unaltered. This status quo arises from the accessibility of all assets in the industrial and field layer networks by assets from other networks. Consequently, there exists no rationale to amend the MAV for these assets within the exploitability vector string.

Moreover, it's important to note that the firewall exclusively protects the assets in the cyber layer, particularly the WebS and VPN. It does not provide additional protection for other assets since the attacker can solely access them from the internal network. Hence, based on this premise, the MAC metric remains unmodified for WS, as well for the industrial and physical layers' assets.

The final step in this phase of our methodology entails the conversion of the modified exploitability score, ranging from 0 to 3.9, into a 1-based scale, thereby relabeling it as the vulnerability score. Table \ref{tab:out} final column showcases the vulnerability score for all identified vulnerabilities within our scenario, representing the output of this particular step. As previously mentioned, the exploitability score has only been modified for the initial four vulnerabilities, which correspond to the assets within the cyber layer network, influenced by the existing topology and security mechanisms.

\begin{table*}
\centering
\caption{Assets, CVEs, original CVSS score, modified exploitability scores, vulnerability score}
\label{tab:out}
\begin{tabular*}{\textwidth}{@{\extracolsep{\fill}}ccccc@{}}
\hline
    \textbf{Asset}  & {CVE} &{CVSS Score}  & {Exp.score} & {Vulnerability Score} \\ \hline
VPN & CVE-2019-11510 & 10 & 2.2 & 0.56 \\ 
WebS & CVE-2017-7269 & 9.8 & 2.2 & 0.56 \\ 
WS & CVE-2017-0143 & 8.1 & 1.6 & 0.41  \\ 
WS & CVE-2017-8692 & 7.5 & 1.2 & 0.31 \\ 
HDB & CVE-2021-1636 & 8.8 & 2.8 & 0.72 \\ 
HDB & CVE-2023-21528 & 7.8 & 1.8 & 0.46 \\
HMI & CVE-2016-5743 & 9.8 & 3.9 & 1\\ 
EWS & CVE-2019-10922 & 9.8 & 3.9 & 1\\
PLC &CVE-2016-9159 & 5.9 & 2.2 & 0.56\\ 
PLC & CVE-2016-8673 & 8.8 & 2.8 & 0.72 \\ \hline
\end{tabular*}
\end{table*}

\subsection{Methodology: Step 2 - Attack Trees}

Step 2 of our methodology involves employing the attack tree framework as detailed in section \ref{AT}. This step involves an analysis of the overall count of vulnerabilities linked to each asset, considering their interdependencies. The utilization of the disjunctive formula (RULE\#3) occurs when the attacker's objective necessitates the exploitation of any single vulnerability. Conversely, for scenarios where the attacker must exploit all vulnerabilities, we employ the conjunctive formula (RULE\#4). We utilise the output (vulnerability score) from step 1 in both disjunctive and conjunctive formulas to assess the comprehensive vulnerability score of each asset. 

In our current scenario, the chosen vulnerabilities represent real-world CVEs. Most of these vulnerabilities, particularly those pertaining to assets in the industrial layer, have been observed within SCADA environments. Our methodology aims to evaluate the interdependencies between vulnerabilities found in the same asset and subsequently refine the overall vulnerability score by applying (RULE\#3) and (RULE\#4). As we describe in Section \ref{AT}, when an adversary exploits two vulnerabilities that individually yield a similar impact, for instance, both vulnerabilities granting direct control upon exploitation, a disjunctive formula should be applied (RULE\#3). This formula accounts for scenarios where the exploitation of either vulnerability independently allows the attack to achieve its objective. Consequently, these vulnerabilities can be regarded as alternative paths to accomplish the same goal.

Conversely, in situations where an attacker needs to exploit multiple vulnerabilities to achieve its objective (RULE\#4), such as first conducting an SQL injection to gain initial access followed by a privilege escalation exploit to attain full control, a conjunctive formula should be employed. This formula represents cases where the exploitation of multiple vulnerabilities is necessary in order for the attacker to achieve their goal.

In our scenario, both the VPN and WebS assets each possess a single vulnerability, eliminating the necessity to employ the attack tree formulas to combine them. However, in contrast, the WS asset contains two vulnerabilities. Each of these vulnerabilities, when successfully exploited, enables remote code execution by the adversary. Therefore, in this instance, we employ the disjunctive formula (RULE\#3).

Among the assets in the industrial layer, both the EWS and HMI showcase single vulnerabilities. Additionally, the HDB asset possesses two vulnerabilities: an SQL injection and a code execution. For demonstrative purposes in our scenario, we assume that both vulnerabilities must be exploited for the attack to attain its goal, so we employ the conjunctive formula (RULE\#4).

In the field layer, the PLC, constituting the ultimate target of the attacker, exhibits two vulnerabilities. While these vulnerabilities do not share a similar impact, we opt to employ the conjunctive formula (RULE\#3), considering that the attacker's objective is to reach and impact the PLC in any adverse manner.

Table \ref{tab:step2} depicts the assets involved in our scenario, detailing the count of vulnerabilities (CVEs) on each asset alongside their interdependencies (RULE\#3 and RULE\#4). Additionally, it showcases the analytical calculations that combine the OR/AND formulas with the vulnerability score obtained from step 1. Finally, the last column of this Table presents the final score, which constitutes the output of this step of our methodology.

\begin{table*}
\centering
\caption{Assets, count of CVEs/asset, interdependencies between CVEs on each asset, formulas' used and final output}
\label{tab:step2}
\begin{tabular*}{\textwidth}{@{\extracolsep{\fill}}ccccc@{}}
\hline
\textbf{Asset} & \textbf{No. of CVE} & \textbf{Interdependencies} & \textbf{Calculations} & \textbf{Step 2 Score} \\ \hline
VPN & 1 & N/A & N/A  & 0.56 \\ 
WebS & 1 & N/A & N/A  & 0.56  \\ 
WS & 2 & OR & 1 - [(1-0.41)*(1-0.31)] & 0.59 \\ 
HDB  & 2 & AND & 0.72 * 0.46 & 0.33 \\ 
HMI  & 1 & N/A & N/A & 1\\ 
EWS  & 1 & N/A & N/A & 1 \\ 
PLC & 2 & OR & 1 - [(1-0.56)*(1-0.72)]  & 0.88 \\ \hline
\end{tabular*}
\end{table*}

During this step, the dynamic vulnerability criticality assessment calculator exhibits the following functionality:

\begin{enumerate}
    \item When multiple vulnerabilities are present in an asset interconnected with a conjunctive relationship, the overall vulnerability score decreases.
    \item Conversely, if multiple vulnerabilities exist in an asset connected by a disjunctive relationship, the overall vulnerability score increases.
    \item In cases where there is at least one vulnerability with the highest score, which in our model at step 2 is one, the vulnerable asset will consistently have the highest vulnerability score, regardless of how many other vulnerabilities exist within this asset, as long as these vulnerabilities are connected with disjunctive relationships.
\end{enumerate}

\subsection{Methodology: Step 3 - FCM}

The conclusive phase, which is the core  of our methodology, entails utilizing the output obtained from step 2 in conjunction with the attack path and the FCM model. This output provides a synthesized vulnerability value that accounts for:
\begin{itemize}
    \item the topology of the environment (step 1),
    \item the deployed security measures (step 1), 
    \item the quantity of vulnerabilities associated with each asset (step 2),  
    \item their interdependencies (step 2).
\end{itemize}

The synthesised vulnerability value acts as a weight within our FCM model, representing the extent of influence between the combined vulnerability score derived from steps 1 and 2, and an asset. To elucidate further, the vulnerability score in our model is derived from the modified exploitability score. Therefore, this value illustrates the degree of exploitability of an asset within the attack path. Furthermore, within our model, the directional arrows signify the causality's direction and nature, depicting the potential pathways through which an adversary can navigate to accomplish their objective.

\begin{figure}
\centering
\includegraphics[width=\columnwidth]{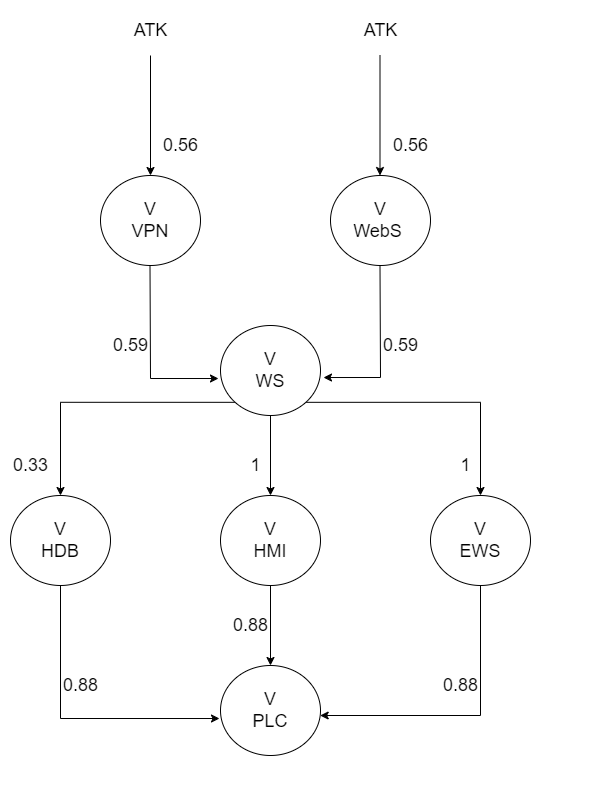}
\caption{FCM model\label{FCM}}
\end{figure}

The construction of our FCM model entails the use of PyCharm, a specialized Integrated Development Environment (IDE) for Python. Additionally, we leverage NetworkX, a Python package designed for the generation, manipulation, and analysis of intricate network structures and dynamics. Figure \ref{FCM} illustrates the  FCM model specifically employed within the context of our scenario.

\subsection{Results}\label{sec:results}

\begin{figure}
  \centering
  \includegraphics[width=\columnwidth]{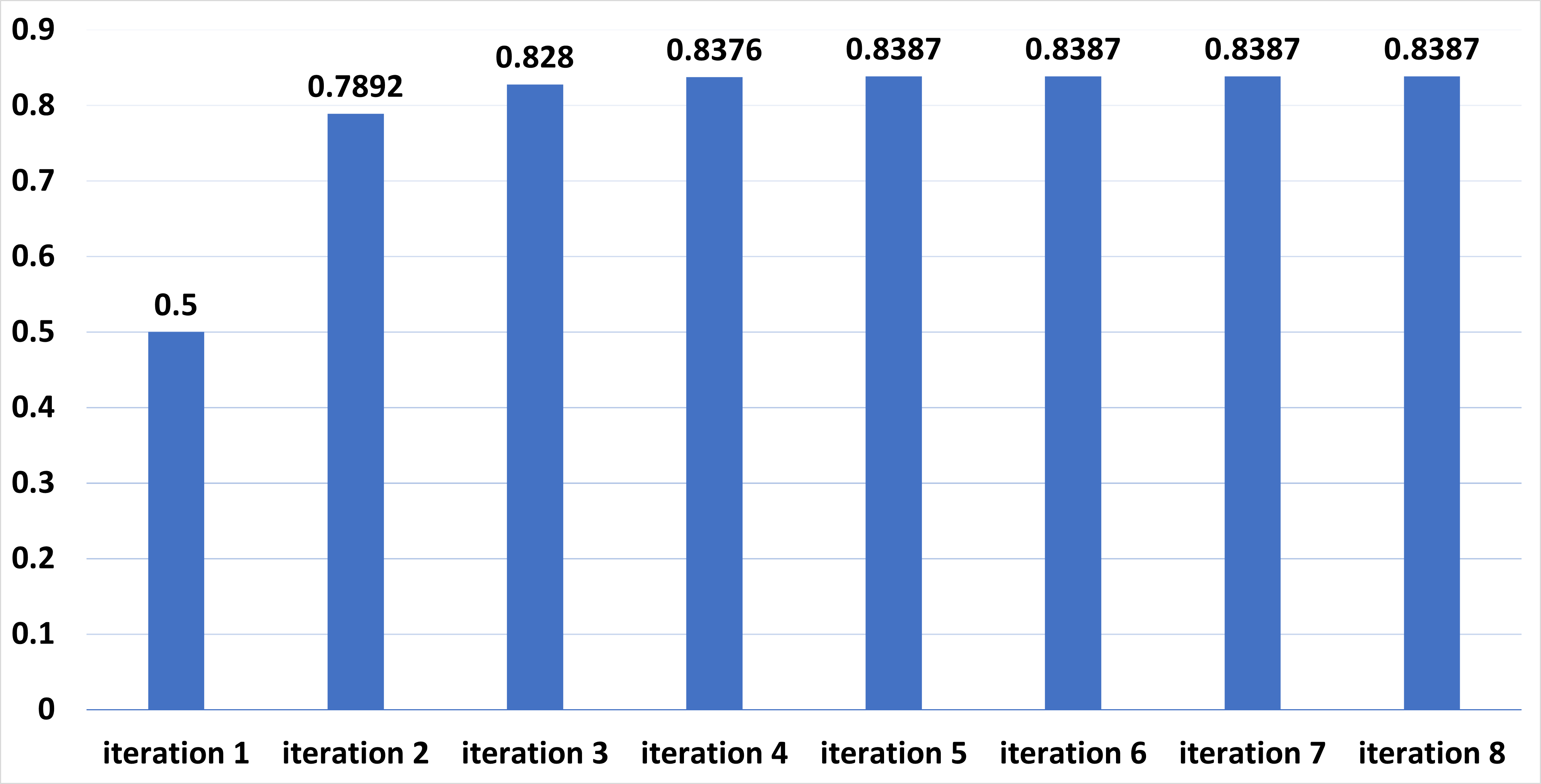}
  \caption{PLC's value}
  \label{ACT}
\end{figure}

Our model reached its equilibrium point after the 5th iteration, resulting in a final score of 0.8387 at the PLC. As outlined earlier, we amalgamated the outputs of step 1 and step 2, utilizing them as weights within our FCM model. Consequently, the final value at the PLC portrays the impact of topology, deployed security mechanisms, the quantity of associated vulnerabilities per asset and their interdependencies, as well as the attack path, along with influence between all assets' vulnerabilities to provide the overall vulnerability score at the PLC. This value is calibrated on a scale of 1, can be linearly scaled to 8.84 within a range of 10 units. Furthermore, this calibrated metric also corresponds to a value of 3.27 within a scale ranging from 0 to 3.9, thereby illustrating the  modified exploitability score of the CVSS vector string at the PLC. As depicted in Figure \ref{ACT}, our system achieved stability (equilibrium point) after undergoing 5 iterations, although the discrepancy between the 4th iteration, with a value of 0.8376, and the 5th iteration, with a value of 0.8387, is minor. The horizontal axis of the graph represents the number of iterations or steps, while the vertical axis represents the value at the PLC per iteration.

\begin{figure}
\centering
\includegraphics[width=\columnwidth]{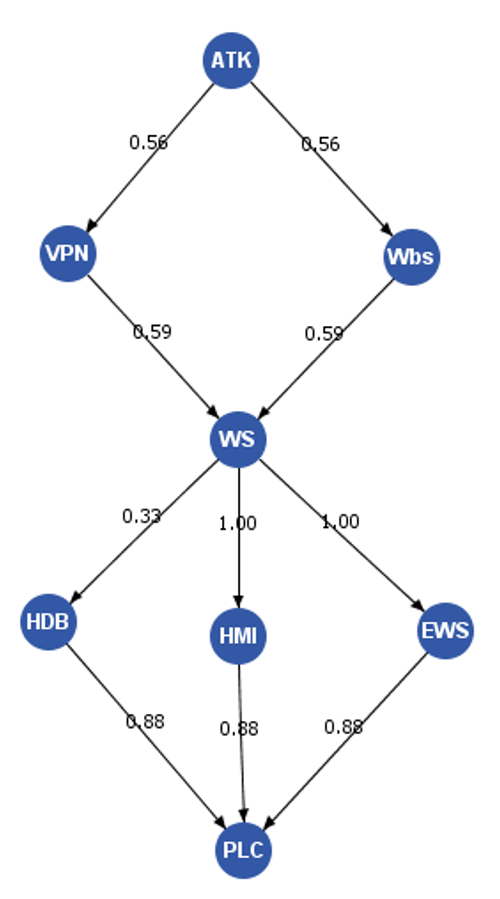}
\caption{FCM model - FCM expert} \label{EnvExp}
\end{figure}

The validity of our FCM model can be affirmed through a comparative analysis with a widely recognized FCM software known as the FCM expert \cite{napoles2018fcm}, \cite{FCMExpert}. Figure \ref{EnvExp} displays the FCM model utilized within the FCM expert software, using the exact same weights and attack path as these that are employed in our model.

Moreover, Table \ref{tab:Expert} showcases the analytical results obtained for all assets in our environment, including the PLC, through the utilization of the FCM expert software. Notably, our model exhibits convergence to an equilibrium point after 5 iterations, yielding a PLC value of 0.8387. Hence, we observed that both our program and the FCM expert software estimated the identical value at the PLC, equating to 0.8387. Additionally, both programs identified the 5th iteration as the equilibrium point. Both models incorporate \textit{Kosko's standard activation rule} \cite{kosko1986fuzzy} and utilize the \textit{sigmoid function as a transfer function}. In the following subsections, we present analogous additional scenarios where we employed our methodology.

\begin{table*}
\centering
\caption{PLC's value - FCM expert}
\label{tab:Expert}
\begin{tabular*}{\textwidth}{@{\extracolsep{\fill}}ccccccccc@{}}
\hline
\textbf{Iterations} & \textbf{ATK} & \textbf{VPN} & \textbf{Wbs} & \textbf{WS} & \textbf{HDB} & \textbf{HMI} & \textbf{EWS} & \textbf{PLC} \\
\hline
1 & 0.5000 & 0.5000 & 0.5000 & 0.5000 & 0.5000 & 0.5000 & 0.5000 & 0.5000 \\
2 & 0.5000 & 0.5695 & 0.5695 & 0.6434 & 0.5412 & 0.6225 & 0.6225 & 0.7892 \\
3 & 0.5000 & 0.5695 & 0.5695 & 0.662 & 0.5529 & 0.6555 & 0.6555 & 0.8280 \\
4 & 0.5000 & 0.5695 & 0.5695 & 0.662 & 0.5544 & 0.6597 & 0.6597 & 0.8376 \\
5 & 0.5000 & 0.5695 & 0.5695 & 0.662 & 0.5544 & 0.6597 & 0.6597 & 0.8387 \\
6 & 0.5000 & 0.5695 & 0.5695 & 0.662 & 0.5544 & 0.6597 & 0.6597 & 0.8387 \\
7 & 0.5000 & 0.5695 & 0.5695 & 0.662 & 0.5544 & 0.6597 & 0.6597 & 0.8387 \\
8 & 0.5000 & 0.5695 & 0.5695 & 0.662 & 0.5544 & 0.6597 & 0.6597 & 0.8387 \\
\hline
\end{tabular*}
\end{table*}

\subsubsection{Scenario A - Additional Firewall}
In the original scenario, a single firewall safeguards the VPN and WebS assets within the enterprise layer network, while a router facilitates the WS connection to assets within the industrial layer network. In this specific scenario (A), we substitute the router with an additional firewall, resulting in the presence of two firewalls for this particular configuration. The original firewall continues to protect the VPN and WS assets, while the additional firewall replaces the router. This substitution should be evaluated according to (RULE\#2) of our methodology, which necessitates an examination of the AC metric in the detected vulnerabilities of the assets within the industrial layer network. By examining the exploitability vector strings of vulnerabilities detected in the assets of the industrial layer network, it becomes evident that their AC metric is (\begin{math}AC:L\end{math}). The introduction of an additional firewall, in conjunction with the (RULE\#2) of our methodology, leads to the transition of the AC metric to (\begin{math}MAC:H\end{math}). This alteration resulted in a change in the modified exploitability score, specifically reducing it. Table \ref{tab:ScA_orig} showcases the original exploitability metrics along with their associated scores, while Table \ref{tab:ScA_mod} illustrates the modified metrics alongside their modified scores.

\begin{table*}
\centering
\caption{Scenario A - Original exploitability metrics and Scores}
\label{tab:ScA_orig}
\begin{tabular*}{\textwidth}{@{\extracolsep{\fill}}cccc@{}}
\hline
\textbf{Asset} & \textbf{CVE} & \textbf{Or. Expl. Metr} & \textbf{Or. Expl. Sc.} \\ \hline
HDB & CVE-2021-1636 & AV:N/AC:L/PR:L/UI:N & 2.8 \\
HDB & CVE-2023-21528 & AV:L/AC:L/PR:L/UI:N & 1.8 \\
HMI & CVE-2016-5743 & AV:N/AC:L/PR:N/UI:N & 3.9 \\
EWS & CVE-2019-10922 & AV:N/AC:L/PR:N/UI:N & 3.9 \\ \hline
\end{tabular*}
\end{table*}

\begin{table*}
\centering
\caption{Scenario b - Modified exploitability metrics and Scores}
\label{tab:ScA_mod}
\begin{tabular*}{\textwidth}{@{\extracolsep{\fill}}cccc@{}}
\hline
\textbf{Asset} & \textbf{CVE} & \textbf{Mod. Expl. Metr.} & \textbf{Mod. Expl. Sc.} \\ \hline
HDB & CVE-2021-1636 & MAV:N/MAC:H/MPR:L/MUI:N & 1.6 \\
HDB & CVE-2023-21528 & MAV:L/MAC:H/MPR:L/MUI:N & 1 \\
HMI & CVE-2016-5743 & MAV:N/MAC:H/MPR:N/MUI:N & 2.2 \\
EWS & CVE-2019-10922 & MAV:N/MAC:H/MPR:N/MUI:N & 2.2 \\ \hline
\end{tabular*}
\end{table*}

To make use of step 2, we convert the aforementioned values into a scale of 1, and rename it as vulnerability score. Table \ref{tab:ScA2} illustrates the associated scores.

\begin{table}[h]
\centering
\caption{Scenario A - Scale-1 scores}
\label{tab:ScA2}
\begin{tabular*}{\columnwidth}{@{\extracolsep{\fill}}cccc@{}}
\hline
\textbf{Asset} & \textbf{CVE} & \textbf{Vulnerability Sc} \\ \hline
HDB & CVE-2021-1636 & 0.41 \\
HDB & CVE-2023-21528 & 0.25 \\
HMI & CVE-2016-5743 & 0.56 \\
EWS & CVE-2019-10922 & 0.56 \\ \hline
\end{tabular*}
\end{table}

Given our selective use of step 2 of our methodology solely for HDB asset, we necessitate the application of a conjunctive formula (RULE\#4)for the CVEs specifically associated with HDB. Consequently, this approach results in three scores: HDB: 0.1, HMI: 0.56, EWS: 0.56.

\begin{figure}
\centering
\includegraphics[width=\columnwidth]{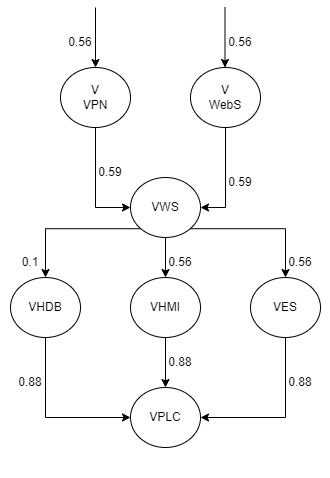}
\caption{Scenario A - FCM model\label{FCMA}}
\end{figure}

The derived scores are integrated into the FCM model as weights, mirroring the scoring method used for the same assets in our original scenario. Consequently, the application of our methodology, leveraging the revised metrics, yielded an overall vulnerability score of 0.8169 at the PLC, graded on a scale of 1. This observation underscores that introducing an additional security mechanism between the aforementioned layers reduced the vulnerability score attributed to the PLC by approximately 2.60\%. Figure \ref{FCMA} presents the FCM model utilised for scenario A.

\subsubsection{Scenario B - Path Elimination}

In the preceding scenarios, both the original setup (involving a router) and the alternative configuration (employing a firewall) present the attacker with three separate pathways to reach their ultimate target, namely the PLC. These are :

\begin{enumerate}
    \item from HDB to PLC with weight equal to 0.88,
    \item from HMI to PLC with weight equal to 0.88,
    \item from EWS to PLC with weight equal to 0.88.
\end{enumerate}

\begin{figure}
\centering
\includegraphics[width=\columnwidth]{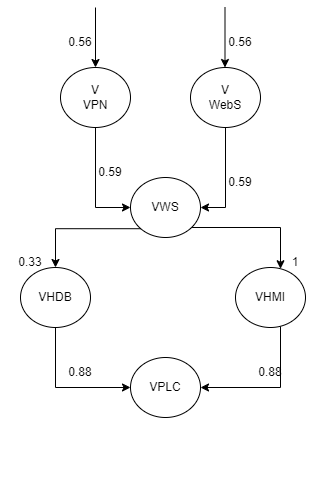}
\caption{Scenario B - FCM model\label{FCMB}}
\end{figure}

Suppose we modify this scenario by eliminating one of the three routes to the PLC. Let us consider that the EWS is no longer a step in the attacker's path to reach the PLC. Consequently, the attacker's access to the PLC is restricted to either HDB or HMI. Despite this modification, the attacker is still able to exploit one of the two CVEs identified within the PLC. Consequently, with the reduction in the number of distinct routes from three to two, each having a weight of 0.88, the overall vulnerability score attributed to the PLC decreases to 0.7442. This reduction signifies an approximate decrease of 11.27\% compared to the original scenario. Figure \ref{FCMB} presents the FCM model utilised for scenario B.

\subsubsection{Scenario C - CVE elimination}
In the context of the original scenario, the adversary possesses the capability to exploit one of the two CVEs interconnected through a disjunctive relationship, identified within the PLC to achieve their objective. Our methodology, as elaborated earlier, attributes a weight equal to 0.88 in this case.

Suppose, however, that instead of having two CVEs linked by a disjunctive relationship, the PLC is affected by only one CVE. Specifically, considering the CVEs and their respective scores extracted from step 2 of our methodology:

\begin{enumerate}
    \item CVE-2016-9159: 0.56
    \item CVE-2016-8673: 0.72
\end{enumerate}

In this scenario, we opt to eliminate one of the two aforementioned CVEs, prioritizing the removal of the more critical vulnerability, i.e., the one with the higher score. Consequently, the PLC is left with a single vulnerability bearing a score of 0.56.

\begin{figure}
\centering
\includegraphics[width=\columnwidth]{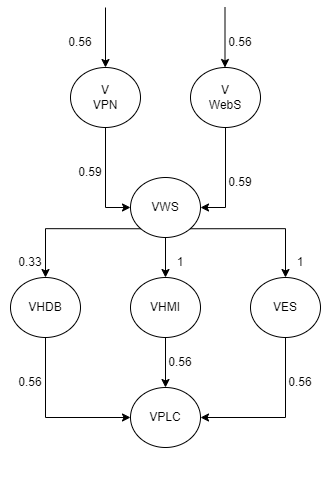}
\caption{Scenario C - FCM model\label{FCMC}}
\end{figure}

The pathways between HDB and PLC, HMI and PLC, and EWS and PLC remain unaltered. However, instead of the previously employed weight of 0.88 between these routes, we utilize the weight associated with the single CVE, which is 0.56. As a result, the PLC's  vulnerability score amounts to 0.7406, reflecting an approximate reduction of 11.77\% compared to the original vulnerability value. Figure \ref{FCMC} presents the FCM model utilised for scenario C.

\subsubsection{Scenario D - A combined approach}

In the final scenario, we amalgamate the security strategies employed in all preceding scenarios. Specifically, we opt to substitute the router between the enterprise layer network and the industrial layer network with a firewall, mirroring the configuration depicted in scenario A. Furthermore, we curtail one of the attacker's pathways to reach the PLC. In this instance, we eliminate the attacker's access to HDB, rendering the HDB asset inconsequential in the attacker's trajectory. Consequently, the attacker's access to the PLC is confined to either EWS or HMI, analogous to scenario B, albeit with a distinct route elimination. 

Lastly, instead of granting the attacker the capability to exploit two vulnerabilities interconnected with a disjunctive relationship, we opt to eliminate one of these vulnerabilities. To be specific, we choose to eradicate the vulnerability with the highest score, aligning with scenario C. Consequently, the PLC is left with only one vulnerability in this refined scenario.

\begin{figure}
\centering
\includegraphics[width=\columnwidth]{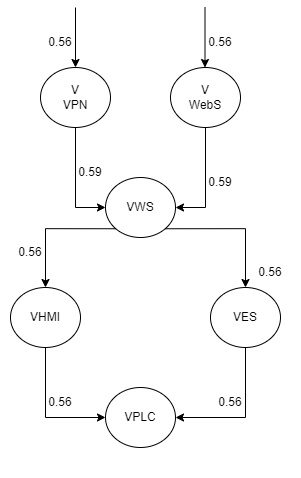}
\caption{Scenario D - FCM model\label{FCMD}}
\end{figure}

In alignment with the preceding scenarios, we integrate all these modifications into our FCM model. The implementation of the various security measures, results in a decrease in the vulnerability score at the PLC, registering at 0.6598. This reduction denotes an approximate decrease of 21.36\%, a substantial improvement compared to the original vulnerability score and the modified scenarios A, B, and C. Figure \ref{FCMD} presents the FCM model utilised for scenario D.

\subsection{Scenarios' Comparison and Discussion}

In the four aforementioned scenarios (A,B,C,D) we implemented three distinct security measures to mitigate the vulnerability value associated with the PLC, culminating in a combined approach in the final scenario. Table \ref{tab:comp} presents a comprehensive summary of the results from the aforementioned scenarios.

Observing the data within this table reveals that the most efficacious standalone security measure was employed in the scenario C, entailing the removal of the one of two CVEs located at the PLC. Both measures employed in scenarios B and C exhibit nearly equal effectiveness in addressing the case study, as evidenced by the marginal difference in the final vulnerability value at the PLC. The combination of security measures, as it was presented in scenario D, emerges as the most effective approach utilized to diminish the vulnerability score at the PLC.

\begin{table}[htbp]
\centering
\caption{Comparison of results from the scenarios}
\label{tab:comp}
\begin{tabular*}{\columnwidth}{@{\extracolsep{\fill}}cccc@{}}
\hline
\textbf{Scenario} & \textbf{Measure} & \textbf{Result} & \textbf{Reduction (\%)} \\
\hline
A & Additional Security Mechanism & 0.8169 & 2.60  \\
B & Path Elimination & 0.7442 & 11.27 \\
C & CVE Elimination & 0.7406 & 11.70 \\
D & Combined Approach & 0.6598 & 21.36\\
\hline
\end{tabular*}
\end{table}

\begin{figure}
\centering
\includegraphics[width=\columnwidth]{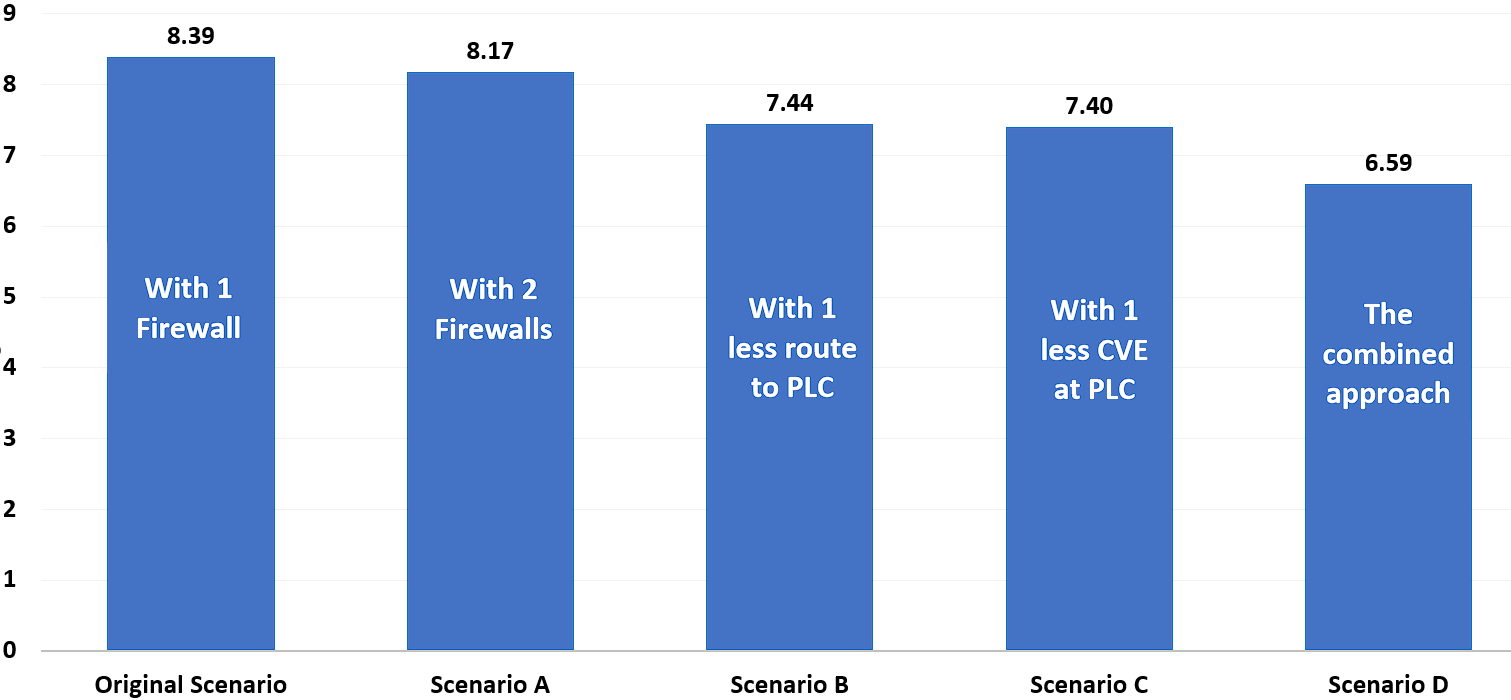} 
\caption{Dynamic Vulnerability Value at PLC}
\label{DVP}
\end{figure}

Figure \ref{DVP} illustrates the comprehensive vulnerability value at the PLC (on a base-10 scale), serving as the conclusive output derived from our methodology. It encapsulates an overall summary across each of the five implemented scenarios, offering a comprehensive overview of our methodology's outcomes.

The substitution of the router between the enterprise layer network and industrial layer network with a firewall results in only a marginal reduction of the vulnerability score at the PLC, amounting to 2.60\%. Consequently, we contend that the installation of an additional firewall between the aforementioned layers proves to be an ineffective security measure. Moreover, the correct configuration of a firewall is a challenging process ~\cite{Uribe2004Automatic}, especially when considering a scenario in which the firewall is implemented within an operational network, accounting for essential adjustments such as the configuration of routes and rules, as well as potential downtime, this security measure may entail more disadvantages than advantages. The implementation of the additional firewall results in the reduction of exploitability scores for the identified vulnerabilities within the assets of the industrial layer network. However, the adversary retains multiple pathways to achieve their objective, namely, negatively affecting the PLC. Consequently, the vulnerability score at the PLC remains nearly unchanged compared to the (original) scenario without the additional firewall. 

A more effective security measure to decrease the vulnerability score at the PLC involves either eliminating potential pathways that adversaries could exploit to reach their target or mitigating the identified vulnerabilities at the PLC. This is exemplified by scenarios B and C, respectively. Considering the context of our methodology applied in an ICS environment, where legacy systems often pose challenges for updates or patches ~\cite{Furnell2014The}, eliminating detected vulnerabilities in the PLC within a real-world environment may prove challenging. Furthermore, it is important to note that HMI or EWS may also be considered legacy systems. Consequently, the aforementioned challenges persist in these components as well. Building on this assumption, it may be prudent to prioritize the elimination of the detected vulnerabilities in other assets that might be simpler or more amenable to updates or patches, such as the HDB. This strategy aims to diminish potential pathways for adversaries, while recognizing the challenges associated with addressing vulnerabilities in legacy systems like the PLC. 

System administrators and security engineers along with SCADA engineers, should evaluate the distinctive cyber-physical topology of their ICS environment ~\cite{Ralston2007Cyber}. Based on this assessment, they can determine which security measures or strategies are more effective to apply without jeopardizing operational stability. Our methodology demonstrates that the introduction of an extra firewall does not have a substantial impact on the vulnerability score at the PLC. Furthermore, we presented that the implementation of the other two security measures
— pathway and CVE elimination — reduce the overall vulnerability score at PLC by 11.27\% and 11.70\% accordingly. In instances where feasible, a combined approach involving multiple security measures should be employed to ensure the maximal reduction in the vulnerability score at the PLC. 

Finally our methodology suggests that security measures applied to the asset-target or to assets directly connected to the asset-target exert a more significant impact on the overall vulnerability score of the asset-target compared to security measures applied to assets located at a greater networking distance from the attacker's perspective.

\section{Conclusions}\label{sec:conc}

In this article, a dynamic vulnerability criticality methodology has been introduced for ICS. This methodology comprises four pivotal steps. Firstly, it involves a pre-process phase aimed at identifying CVEs and extracting exploitability-only metrics from their CVSS vector strings. Subsequently, AV and AC metrics are adjusted based on the environment's topology and applied security mechanisms, to compute the modified exploitability scores. Thirdly, it utilizes attack-tree formulas to assess the quantity of vulnerabilities within each asset and their interdependencies. Finally, it integrates the results from the previous steps along with the attack path into the FCM model to generate a dynamic vulnerability score.

The effectiveness of this approach was demonstrated through a comprehensive case study highlighting the practical application and efficacy of our proposed methodology. In addition to the initial case study that was presented and examined, this study introduces and analyzes four supplementary scenarios. These scenarios are designed to specifically address the reduction of the vulnerability score at the PLC, representing the ultimate target for potential attackers. Each scenario (A,B,C) implements distinct security measures, and their outcomes are systematically compared with the results obtained from the original case study. Furthermore, a combined security approach is introduced to demonstrate its effectiveness in reducing the vulnerability score at PLC. Moreover, the study delves into a discussion of the findings, taking into account the unique characteristics inherent in an ICS environment. The aim is to provide valuable insights and recommendations for security and SCADA engineers to enhance the robustness of their systems.

In our future endeavors, the primary objective is to integrate our model into a dynamic risk scheme, enabling the provision of dynamic vulnerability values alongside dynamic threat probability assessments and impact estimations. Therefore, our forthcoming work will concentrate on devising a dynamic probability of threat occurrence calculator that will complement our established model. This dynamic, and even proactive, threat probability calculator, when combined with our existing model and the impact values (static or even dynamic), aims to yield an authentic dynamic risk score that will facilitate proactive situational awareness. 

\section*{Declarations}
\textbf{Funding}: This work was partially funded by the Horizon Europe program through the projects “ Reliability, Resilience and Defense Technology for the Grid ” (R2D2) (Grant agreement ID: 101075714).

\textbf{Competing interests}: The authors have no competing interests to declare that are relevant to the content of this article.

\textbf{Ethical approval}: This manuscript does not contain any studies with human participants or animals performed by any of the authors.

\textbf{Research data}: The data utilized for the purpose of this paper was obtained from the NIST NVD website, publicly accessible.

\appendix

\section{Abbreviations}
The following abbreviations are used in this manuscript:\\

\begin{table}[ht]
\begin{tabular}{@{}ll}
AC & Attack Complexity \\
AR & Availability Requirement \\
AV & Attack Vector \\
CAPEC & Common Attack Pattern Enumeration and Classification \\
CIA & Confidentiality, Integrity, and Availability \\
CR & Confidentiality Requirement \\
CVE & Common Vulnerabilities and Exposures \\
CVSS & Common Vulnerability Scoring System \\
DRA & Dynamic Risk Assessment \\
EMFM & Extended Multilevel Flow Model \\
EWS & Engineering WorkStation \\
FCM & Fuzzy Cognitive Map \\
HDB & Historical Databases \\
HMI & Human Machine Interface \\
ICS & Industrial Control System \\
ICT & Information and Communication Technologies \\
IDS & Intrusion Detection System \\
IR & Integrity Requirement \\
MA & Modified Availability \\
MAC & Modified Attack Complexity \\
MAV & Modified Attack Vector \\
MC & Modified Confidentiality \\
MI & Modified Integrity \\
MPR & Modified Privileges Required \\
MS & Modified Scope \\
MUI & Modified User Interaction \\
NIST & National Institute of Standards and Technology \\
NVD & National Vulnerability Database \\
OWL & Web Ontology Language \\
PLC & Programmable Logic Controller \\
PR & Privileges Required \\
S & Scope \\
SCADA & Supervisory Control and Data Acquisition \\
SWRL & Semantic Web Rule Language \\
UI & User Interaction \\
VPN & VPN server \\
WebS & Web Server \\
WS & Administrative WorkStations \\
\end{tabular}
\end{table}

\bibliographystyle{elsarticle-num} 
\bibliography{refers}

\end{document}